%% file: main.tex
\documentclass[conference]{IEEEtran}

\usepackage{amsmath, amssymb}
\usepackage[noend]{algpseudocode}  
\usepackage{algorithm}
\usepackage{algpseudocode}

\usepackage{xcolor}   
\usepackage{times}
\usepackage{comment}
\usepackage{tcolorbox} 
\usepackage{xspace} 
\usepackage{cite}
\usepackage{graphicx}
\usepackage{caption}

\usepackage{booktabs}    
\usepackage{array}       
\usepackage{multirow}    
\usepackage{enumitem}
\usepackage{makecell}
\usepackage{flushend}

\usepackage{fancyhdr}
\fancypagestyle{header}{
  \fancyhf{} 
  \fancyhead[C]{\footnotesize The paper was published on IEEE INFOCOM 2026.} 
}

\usepackage{url}
\usepackage{hyperref}    

\newcommand{\pname}{\texttt{EExAPP}\xspace}
\newcommand{\eeactor}{\texttt{EE Actor}\xspace}
\newcommand{\rsactor}{\texttt{RS Actor}\xspace}
\newcommand{\hz}[1]{\textbf{\textcolor{red}{[#1]}}}
\newcommand{\cmark}{\ding{51}}  
\newcommand{\xmark}{\ding{55}}

\begin{document}
%
\title{EExApp: GNN-Based Reinforcement Learning for Radio Unit Energy Optimization in 5G O-RAN}


\author{
    \IEEEauthorblockN{Jie Lu, Peihao Yan, and Huacheng Zeng}
    
    \IEEEauthorblockA{{Department of Computer Science and Engineering, Michigan State University, USA}} 
    
}

\maketitle

\thispagestyle{header}

\begin{abstract}

With over 3.5 million 5G base stations deployed globally, their collective energy consumption (projected to exceed 131 TWh annually) raises significant concerns over both operational costs  and environmental impacts.
In this paper, we present \pname, a deep reinforcement learning (DRL)-based xApp for 5G Open Radio Access Network (O-RAN) that jointly optimizes radio unit (RU) sleep scheduling and distributed unit (DU) resource slicing. 
\pname uses a dual-actor-dual-critic Proximal Policy Optimization (PPO) architecture, with dedicated actor-critic pairs targeting energy efficiency and quality-of-service (QoS) compliance. A transformer-based encoder enables scalable handling of variable user equipment (UE) populations by encoding all-UE observations into fixed-dimensional representations. To coordinate the two optimization objectives, a bipartite Graph Attention Network (GAT) is used to modulate actor updates based on both critic outputs, enabling adaptive trade-offs between power savings and QoS.
We have implemented \pname and deployed it on a real-world 5G O-RAN testbed with live traffic, commercial RU and smartphones. Extensive over-the-air experiments and ablation studies confirm that \pname significantly outperforms existing methods in reducing the energy consumption of RU while maintaining QoS. 
The source code is available at \cite{github}.


\end{abstract}


\IEEEpeerreviewmaketitle

\section{Introduction}
\input{1_Intro}


\section{System Modeling}

\input{3_Problem_Formulation}
\label{PF}

\section{EExApp: Design}

\input{4_Methodology}

\label{4_Methodology}

\section{Experimental Evaluation}

\input{5_Experiments}

\label{5_Experiments}

\section{Related Work}

\input{2_Related_Work}

\section{Conclusion}

\input{6_Conclusion}
\label{6_Conclusion}


\section*{Acknowledgment}
This work was supported in part by NTIA PWSCIF Award 26-60-IF010 and NSF Grant CNS-2312448.

\clearpage

\bibliographystyle{IEEEtran}
\bibliography{references.bib}

\end{document}

%% file: 1_Intro.tex
Over 3.5 million 5G base stations (BSs) have been deployed globally, and the number continues to grow \cite{han2020energy}.
The total electricity consumption of the 5G radio access network (RAN) is projected to exceed 131 terawatt-hours per year \cite{huttunen2023base}, raising serious concerns not only about operational expenses but also environmental impact due to carbon emissions.
Among various components of the 5G RAN, radio units (RUs) are the most energy-intensive, accounting for up to 80\% of the total energy consumption \cite{gsm2022mobile}. This is primarily due to their essential role in managing the physical (PHY) layer of the radio interface, including signal transmission and reception, low-PHY signal processing, and other critical operations that directly influence energy consumption \cite{kundu2025toward}.
As 5G networks continue to expand and adopt wider bandwidths, the energy demand of individual RUs continues to grow, making them a key focus for energy efficiency optimization within the RAN architecture.


\begin{figure}
    \centering
    \includegraphics[width=\linewidth]{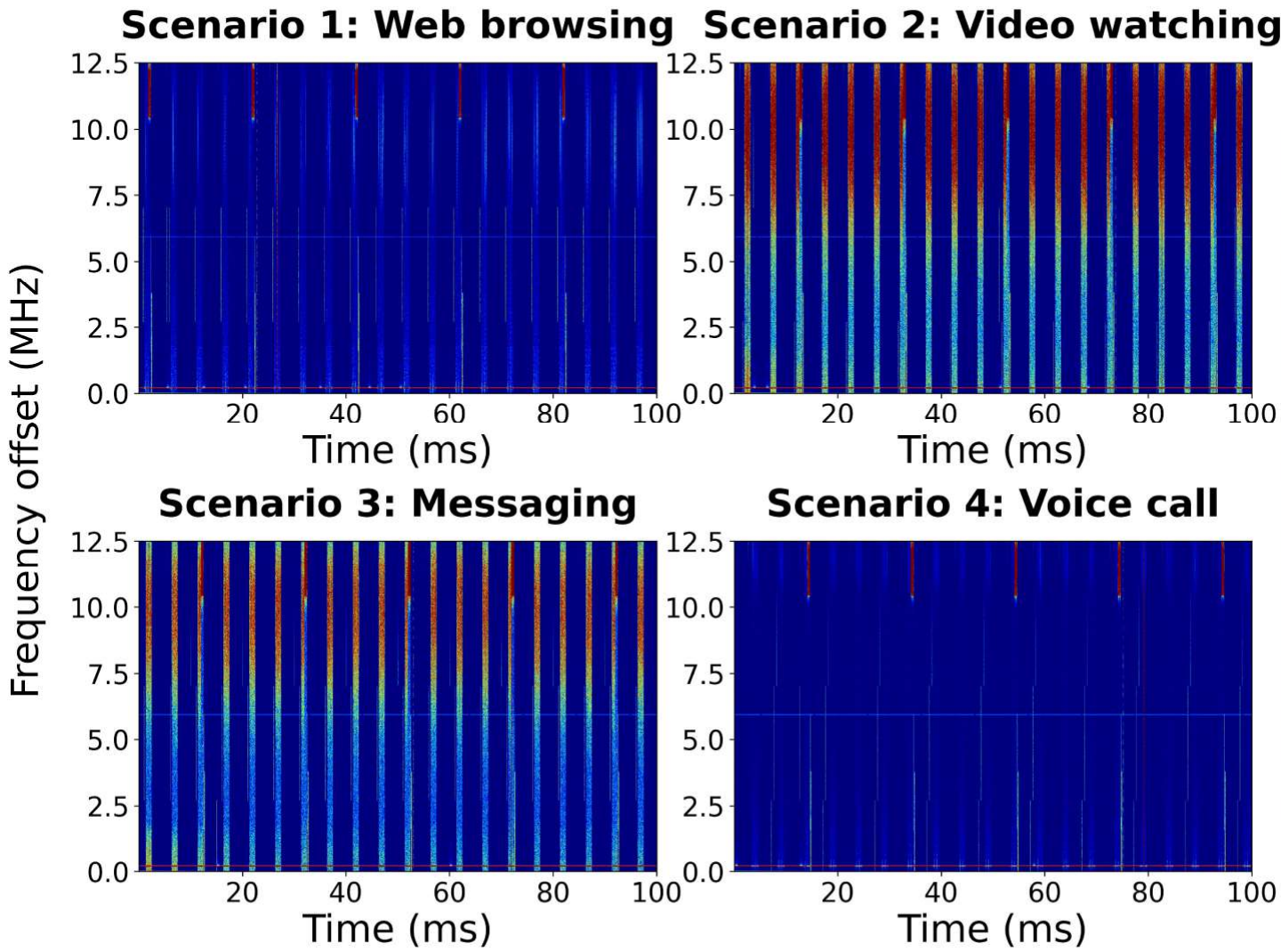} 
    \caption{The radio transmission pattern instances of a 5G base station.}
    \label{fig:motivation}
\end{figure}


Measurements of real-world 5G BSs show that their transmissions are highly bursty, depending on upper-layer data traffic patterns \cite{lozano2025kairos}.
Fig.~\ref{fig:motivation} shows our measurements of an O-RAN system serving smartphones engaged in typical activities such as web browsing, video streaming, messaging, and voice calls.
The observed transmission pattern reveals a comb-like structure, where bursts of radio activity are interleaved with idle time slots.
These idle slots, which dominate the millisecond-level timeline, align with findings in \cite{lozano2025kairos} showing that inactive periods account for more than half of the operating time in median cells.
This presents a significant opportunity for energy savings by transitioning RUs into sleep mode during short idle intervals, without compromising service quality.
In practice, RU sleep scheduling is closely tied to resource slicing at the distributed unit (DU) of the RAN.
A joint optimization of these two functions can maximize RU sleep opportunities while meeting the quality-of-service (QoS) requirements of user equipment (UE), thereby balancing energy efficiency with network performance.
Furthermore, the recent adoption of the O-RAN architecture provides a unique opportunity to implement this joint optimization using xApps deployed within the Near-Real-Time RAN Intelligent Controller (Near-RT RIC), enabling near-real-time adaptation to dynamic network conditions \cite{polese2023understanding, yan2025near, yan2025xdiff}.

In this paper, we propose a joint optimization framework for RU's sleep scheduling and DU's resource slicing in a 5G O-RAN system, with the objective of maximizing the energy efficiency of RU while maximally meeting the QoS demands of UEs.
To attain this objective, we introduce \pname, a deep reinforcement learning (DRL)-based solution implemented as an xApp within the Near-RT RIC.
\pname employs a \textit{dual-actor-dual-critic} architecture built upon the model-free Proximal Policy Optimization (PPO) algorithm.
By taking the representation of the network conditions, \pname includes two actor networks for policy generation:
(i) \eeactor generates discrete sleep scheduling decisions, determining the sleep pattern of the RU per signal frame;
(ii) \rsactor produces continuous resource slicing decisions for DU to ensure the QoS compliance.
Each actor is paired with a dedicated critic network, which independently evaluate energy efficiency and QoS performance.
The two actors are trained using separate update strategies, allowing the framework to optimize distinct objectives effectively.
Compared to its single-actor-single-critic (SASC) counterpart, this dual-actor-dual-critic architecture accelerates the learning convergence and enhances the adaptability of the DRL model.


One challenge in designing \pname is the time-varying number of UEs in cellular networks. In real-world networks, the UE population fluctuates as devices enter, exit, or hand over between cells \cite{yan2025xdiff}.
Since the DRL framework relies on all-UE observations (e.g., traffic demand and performance metrics), the input state space of \pname varies over time, violating the fixed input dimensionality required by neural networks.
To address this, \pname incorporates a Transformer-based encoder \cite{vaswani2017attention}, designed in accordance with the O-RAN standard.
This module processes a variable-length set of key performance indicators (KPIs) collected from the RAN's central unit (CU) and DU, encoding them into a fixed-dimensional latent representation.
The encoder leverages self-attention to capture contextual relationships among UEs and supports scalability to varying input sizes.
This approach significantly reduces the parameter count and inference complexity of the DRL model while preserving capacity to learn effective policies.

Another challenge lies in modeling the interaction between the two actor-critic pairs. In O-RAN, sleep scheduling and resource slicing pursue distinct objectives: one for energy savings and the other for QoS satisfactions.
Training these modules in isolation can lead to suboptimal coordination. For example, extending the sleep duration of RU may degrade QoS, whereas strict QoS enforcement may limit energy-saving opportunities.
To capture these inter-dependencies, \pname employs a bipartite Graph Attention Network (GAT) to connect the two actor-critic pairs.
The bipartite GAT dynamically learns how much influence each critic should exert on each actor during training.
This allows each actor to receive a weighted combination of both critic values, effectively balancing the two objectives while maintaining modular learning for each actor-critic pair.

We implemented \pname as an xApp and deployed it on a 5G O-RAN testbed consisting of one commercial BS and eight smartphones.
Experimental results show that the dual-actor–dual-critic architecture consistently outperforms its single-actor–single-critic counterpart.
Ablation studies further confirm that both the encoder and GAT modules significantly improve \pname's policy learning.
End-to-end system evaluations demonstrate that \pname achieves superior performance compared to state-of-the-art (SOTA) baselines. This work advances the SOTA as follows:
\begin{itemize}
\item
\pname presents a novel joint optimization framework for RU sleep scheduling and DU resource slicing in 5G O-RAN, using a dual-actor-dual-critic DRL architecture that efficiently handles dynamic UE populations.

\item
\pname is deployable in commercial O-RAN systems, with decision-making offloaded to the Near-RT RIC, which operates under more relaxed latency constraints compared to the DU.

\item
Extensive over-the-air experiments show that \pname significantly outperforms the existing approaches.
\end{itemize}

%% file: 3_Problem_Formulation.tex
\subsection{A Primer on 5G NR and O-RAN}

\textbf{5G and Beyond.}
5G New Radio (NR) introduces a flexible and highly scalable frame structure designed to accommodate the diverse service requirements of 5G networks, ranging from enhanced mobile broadband (eMBB) and ultra-reliable low-latency communications (URLLC) to massive machine-type communications (mMTC) \cite{shafi20175g}. The 5G NR frame structure is built around a 10-ms frame, which is divided into subframes, each lasting 1 ms. These subframes are further divided into slots, and each slot can consist of 14 symbols. The flexibility of the frame structure is largely driven by numerology, which determines the time-frequency grid's granularity based on different subcarrier spacings.
While cellular networks evolve from 5G to 6G, the frame structure and numerology are likely to remain largely the same to ensure backward compatibility.

\textbf{O-RAN.}
The architecture of O-RAN is designed to be open, flexible and modular, enabling the integration of diverse vendors' equipment and improving network management, orchestration, and performance \cite{santos2025managing}. 
RU, DU, and CU are the key components in this architecture. RU is responsible for handling the lower layers of the radio interface, including signal processing for PHY transmission and reception. It manages the antenna and RF components, which are physically located at the cell sites. DU is responsible for the lower layers of the protocol stack (L2 and L3), including the MAC, RLC, and PDCP layers \cite{brik2024explainable}. It performs functions like scheduling, beamforming, and other radio resource management tasks. CU is in charge of the upper layers of the protocol stack, such as SDAP, RRC, and NAS. 
The O-RAN architecture also includes the Near-RT RIC, which plays a central role in enhancing RAN's performance through near-real-time decision-making and optimization. 
It can interact with the DU and CU to monitor the network performance and adjust parameters. The Near-RT RIC operates in coordination with xApps, which are specific applications deployed within the Near-RT RIC to provide customized functionalities such as resource management, traffic steering, and energy efficiency optimization.

\textbf{Energy Saving in 3GPP.}
3GPP has standardized a range of energy-saving mechanisms and sleep modes aimed at reducing the power consumption of RANs. These mechanisms target both network-side components (e.g., RU and DU) and UEs. At the network level, Carrier-Level Sleep allows dynamic activation and deactivation of carriers based on traffic demand, effectively enabling small cells or secondary carriers to power down during low-utilization periods \cite{3gpp38300, 3gpp38401}. Transmitter/Receiver (TRx) On/Off, also referred to as RU Sleep, selectively powers down RF chains or antenna elements within a radio unit when transmission or reception is not needed \cite{3gpp38104,3gpp38811,3gpp38300}. Similarly, MIMO Layer/Chain Deactivation allows high-order MIMO systems to reduce energy consumption by disabling unused antenna paths \cite{3gpp38104}.
In addition, a mechanism referred to as DU Sleep Mode Control has been defined in 3GPP \cite{3gpp38401}, allowing CU to initiate the sleep mode of DU. 
On the UE side, 3GPP defines Discontinuous Reception (DRX) in both LTE and NR to conserve battery power during idle periods, which also indirectly contributes to overall RAN energy savings.

\begin{table}
\renewcommand{\arraystretch}{1.0}
\caption{Notation.}
\label{tab:notation}
\centering
\begin{tabular}{ll}
\toprule
\textbf{Symbol} & \textbf{Explanation} \\
\midrule
$\mathcal{I}$  & The set of slices \\
$\mathcal{K}_i$ & The set of UEs in slice $i \in \mathcal{I}$ \\
$q_{t,k}$    &   Average data throughput of UE $k$ in timestep $t$ \\
$d_{t,k}$    &   Average data queueing delay of UE $k$ in timestep $t$ \\
$\mathbf{s}_t$ &  Observation of RAN at time frame $t$ \\
$\mathbf{\hat{s}}_t$ &  Encoded RAN observation, i.e., state of RL, in timestep $t$ \\
$V(\mathbf{\hat{s}}_t)$    &  Individual value function of critic in RL \\
$\hat{V}(\mathbf{\hat{s}}_t)$    &  Aggregated value functions of critic in RL \\
$\boldsymbol{\alpha}_t$  &  $\boldsymbol{\alpha}_t = [a_t, b_t, c_t]$ is for RU sleep control (RL's sleep action)    \\ 
            & in time frame $t$, where $b_t$ is the \# of time slots for sleep\\
$\boldsymbol{\beta}_t$  &  $\boldsymbol{\beta}_t = [\beta_{t,i}]$ is for RU's slicing control (RL's slicing action)    \\ 
\bottomrule
\end{tabular}
\end{table}


\subsection{Problem Formulation}

We formulate the joint energy saving and resource slicing problem as an optimization problem. 
Table~\ref{tab:notation} lists our key symbol representations.



\begin{figure}
    \centering
    \includegraphics[width=\linewidth]{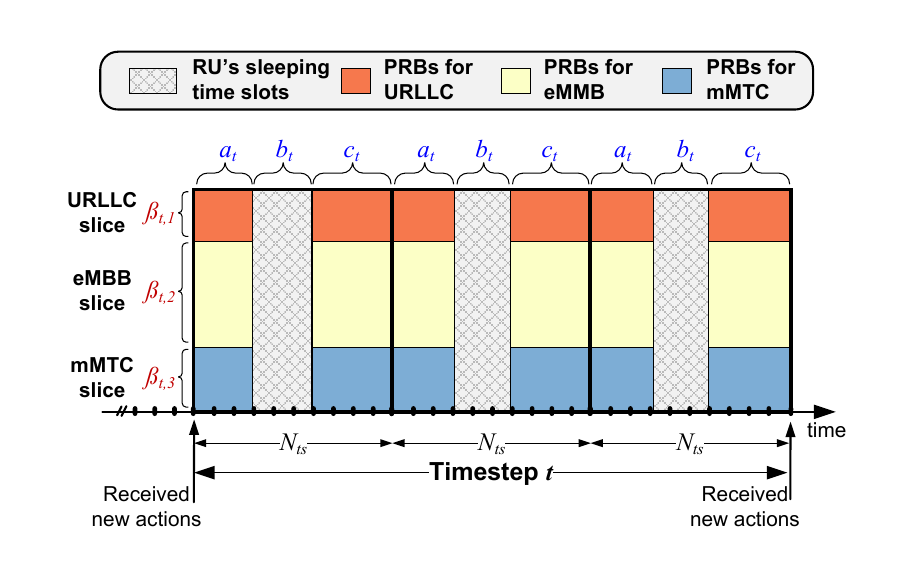} 
    \caption{Illustrating the online policy for joint resource slicing and power-saving optimization at an O-RAN RU.}
    \label{fig:policy_illustration_v1}
\end{figure}

\textbf{Resource Slicing:}
Consider a 5G O-RAN system that comprises RU, DU, CU and Near-RT RIC. 
Denote $\mathcal{I}$ as the set of slices in the RAN, with $I = |\mathcal{I}|$.
Denote $\mathcal{K}_i$ as the set of UEs assigned within slice $i \in \mathcal{I}$, with $\mathcal{K} = \cup_{i\in \mathcal{I}} \mathcal{K}_i$.
Referring to Fig.~\ref{fig:policy_illustration_v1}, denote $\beta_{t,i}$ as the percentage of the physical resource blocks (PRBs) that are allocated for slice $i \in \mathcal{I}$ in timestep $t$. 
Denote $\boldsymbol{\beta}_t = [\beta_{t,1}, \beta_{t,2}, \ldots, \beta_{t,i}]$, with $\sum_{i \in \mathcal{I}} \beta_{t,i} = 1$. 
$\boldsymbol{\beta}_t$ is the online optimization variables for resource slicing in $t$.

\textbf{Sleep Scheduling:}
Again, referring to Fig.~\ref{fig:policy_illustration_v1}, each frame is 10 ms and divided into ${N}_\text{ts} =2^{\mu}\times10$ slots, depending on the 3GPP numerology adopted by the O-RAN, with $\mu \in \{0,1,2,3,4\}$ \cite{3gppR11716574}.
Among the ${N}_\text{ts}$ slots in a frame, denote $b_t$ as the number of slots scheduled for the  sleeping of the RU in timestep $t$. 
To denote the position of sleep slots, we let $a_t$ and $c_t$ be the number of active slots before and after the sleep period. 
Then, we have $a_t + b_t + c_t = N_\text{ts}$.
For simplicity, we let $\boldsymbol{\alpha}_t = [a_t, b_t, c_t]$. 
$\boldsymbol{\alpha}_t$ is the online optimization variables for the sleep scheduling of RU.
We note that, as shown in Fig.~\ref{fig:policy_illustration_v1}, an timestep may span over multiple frames.
In this case, we apply the same sleep scheduling decision to individual frames. 

\textbf{QoS Demands:}
Different slices are to meet different QoS demands. 
Denote $Q_i$ and $D_i$ as the target throughput and delay demands of UEs in slice $i \in \mathcal{I}$. 
Denote $q_{t,k}$ as the achievable throughput and $d_{t,k}$ as the achievable delay of UE $k \in \mathcal{K}$ in timestep $t$. 
Obviously, many factors may affect $q_{t,k}$ and $d_{t,k}$, including the variables of resource slicing and sleep scheduling. 
We denote them as: 
$q_{t,k}=f(\boldsymbol{\alpha}_t,\boldsymbol{\beta}_t,\boldsymbol{s}_t, k)$
and 
$d_{t,k}=g(\boldsymbol{\alpha}_t,\boldsymbol{\beta}_t,\boldsymbol{s}_t, k)$, 
where $\boldsymbol{s}_t$ is the global RAN state/condition in timestep $t$.
Then, the QoS constraints can be expressed as:
$q_{t,k} \ge Q_i$
and 
$d_{t,k} \le D_i$ 
for $k \in \mathcal{K}_i$ and $i \in \mathcal{I}$. 
It is worth noting that in practice, $f(\cdot)$ and $g(\cdot)$ are unknown and hard to estimate due to the complex networks.

\textbf{Formulation:}
The objective of this work is to develop a policy that can maximize the sleep time of the RU while meeting the QoS demands of individual UEs in the network. 
Specifically, we model the optimization objective by $\max(\frac{b_t}{N_\text{ts}})$.
Based on this objective, we formulate the optimization as follows:
\begin{subequations} 
\label{eq:constraints}
\begin{align}
&\max \quad   \mathbb{E} \left[ \lim_{T \to \infty} \frac{1}{T}\sum_{t=1}^{T} \frac{b_t}{N_\mathrm{ts}} \right] \label{objective} \\
\text{s.t.} \; & q_{t,k}=f(\boldsymbol{\alpha}_t,\boldsymbol{\beta}_t,\boldsymbol{s}_t, k) \ge {{Q}_i}, \quad  k\in \mathcal{K}_i, i \in \mathcal{I},    \label{ea} \\
& d_{t,k}=g(\boldsymbol{\alpha}_t,\boldsymbol{\beta}_t,\boldsymbol{s}_t, k) \leq {{D}_i}, \quad   k\in \mathcal{K}_i, i \in \mathcal{I},  \label{eb} \\
& a_t+b_t+c_t = {N}_\mathrm{ts},  \label{ec} \\
& \sum_{i \in \mathcal{I}} \beta_{t,i} = 1, \label{ed}
\end{align}
\label{eq:opt}
\end{subequations}
\!\!\!where $t= 1, 2, \ldots$ is the timestep of decision making process. 
${Q}_i$ and ${D}_i$ are the throughput and delay demands of slice $i$, respectively. 
Constraints (\ref{ea}) and (\ref{eb}) ensure that the QoS demands are met for every UE in every slice.
(\ref{ec}) and (\ref{ed}) characterizes the underlying relations of optimization variables, as explained before.

\subsection{MDP Modeling and Relaxation}

The problem in \eqref{eq:opt} is a constrained stochastic optimization problem with mixed-integer variables. Due to the time-varying network condition $\boldsymbol{s}_t$, the behavior of the system $q_{t,k}$ and $d_{t,k}$ evolve dynamically over time with the complex yet unknown functions $f(\cdot)$ and $g(\cdot)$. This non-stationary and black-box nature makes the problem impractical to solve using static and deterministic methods. 
DRL has emerged as an efficient approach for stochastic optimization, as it can learn from the dynamic environment and adapt its policy based on observed states. In particular, model-free DRL is well-suited for online decision-making in this scenario, as it operates effectively without requiring explicit knowledge of the underlying functions $f(\cdot)$ and $g(\cdot)$ \cite{polese2023understanding}.

In what follows, we formulate the optimization problem as a Markov Decision Process (MDP) in the O-RAN environment, aimed at developing a DRL-based xApp for online optimization of joint sleep scheduling and resource slicing. 

\begin{table}[!t]
    \caption{The per-UE KPI observations from the RAN.}
    \label{tab:kpi_mac_data}
    \centering
    \footnotesize
    \setlength{\tabcolsep}{1.5pt}
    \renewcommand{\arraystretch}{1.05}
    \begin{tabular}{p{1.5cm} p{2.0cm} p{1.0cm} p{3.9cm}}
        \toprule
        \textbf{Data Type} & \textbf{Metric Name} & \textbf{UL/DL} & \textbf{Description} \\
        \midrule
        \multirow{4}{*}{KPM data}
            & PDCP SDU & UL\&DL & Data payload unit \\
            & Delay & DL & SDU delay of RLC  \\
            & Throughput & UL\&DL & Actual data achieved by UE \\
            & PRB & UL\&DL & Physical resource blocks \\
        \midrule
            \multirow{8}{*}{MAC data} 
            & TBS & DL & Current transport block size \\
            & RB & DL & Scheduled resource blocks \\
            & PUSCH SNR & UL & SNR of PUSCH \\
            & PUCCH SNR & UL & SNR of PUCCH \\
            & CQI & DL & Channel quality indicator \\
            & MCS & UL\&DL & Modulation and coding scheme \\
            & PHR & UL & Power headroom report \\
            & BLER & UL\&DL & Block error rate \\
        \bottomrule
    \end{tabular}
\end{table}

\begin{itemize}[leftmargin=0.15in]
    \item \textbf{State $\boldsymbol{s}_t$:} 
    The state representation in our DRL model is constructed from key performance indicator (KPI) generated by the operation of the RAN, including both CU and DU. This data captures the performance of individual UEs and overall network conditions. 
    Table~\ref{tab:kpi_mac_data} summarizes the KPI data that an xApp can obtain from the RAN via standard E2 interface. It includes UE-level metrics such as throughput, delay, and allocated physical resource blocks (PRBs), as well as MAC-layer indicators like the signal-to-noise ratio (SNR) of the Physical Uplink Shared Channel (PUSCH) and Physical Uplink Control Channel (PUCCH), power headroom report (PHR), modulation and coding scheme (MCS). 
    \textit{We stress that, while there is a considerable body of work on DRL for 5G network optimization, most existing studies rely on simulated data. In contrast, this work uniquely models the DRL framework using a realistic dataset.}

\item \textbf{Action $\boldsymbol{a}_t = [ \boldsymbol{\alpha}_t,\boldsymbol{\beta}_t ]$:}
As illustrated in Fig.~\ref{fig:policy_illustration_v1}, the action in our DRL model comprises two components:
$\boldsymbol{\alpha}_t=[a_t,b_t,c_t]$, representing discrete decisions for sleep scheduling, and
$\boldsymbol{\beta}_t = [\beta_{t,1}, \beta_{t,2}, \ldots, \beta_{t,I}]$, representing continuous decisions for resource slicing.
This defines a hybrid action space. The discrete action $\boldsymbol{\alpha}_t$ determines the time scheduling of active and sleep slots for the RU, while the continuous action $\boldsymbol{\beta}_t$ controls the resource allocation among multiple network slices to support key 5G service types such as URLLC, eMBB, and mMTC.

\item \textbf{Reward $r_t$:} 
\textit{Problem \eqref{eq:opt}} aims to maximize the energy efficiency of RU while ensuring that each service slice satisfies QoS constraints. However, in practice, due to the unpredictable traffic patterns and limited resources in the wireless channel, the QoS constraints may not always be strictly satisfied. To address this, we reformulate \textit{Problem \eqref{eq:opt}} using the Lagrangian relaxation technique, which allows for soft constraint violation by introducing penalty terms into the objective function. By incorporating the QoS constraints into the objective function, we have
\begin{align}
    L(\pi, \lambda_p, \lambda_d) =\; 
    & \mathbb{E} \Bigg[ \frac{1}{T} \sum_{t=1}^{T} \Bigg(\frac{b_t}{N_\mathrm{ts}} 
    - \lambda_q \sum_{i \in \mathcal{I}} \sum_{k \in \mathcal{K}_i} ( 1-\frac{q_{t,k}}{{Q}_i})^+ 
    \nonumber \\
    & \quad\quad - \lambda_d \sum_{i \in \mathcal{I}} \sum_{k \in \mathcal{K}_i}  (\frac{{d}_{t,k}}{{D}_i}-1)^+ \Bigg) \Bigg],
    \label{eq:lagrangian}
\end{align}
where $(\cdot)^+ = \max(0, \cdot)$, 
$\lambda_q$ and $\lambda_d$ are Lagrangian multipliers for the throughput and delay constraints, respectively.
In our experiments, we observed that some packets may have large delay and thus dominate the objective function. 
To address this issue, we define 
$\hat{d}_{t,k} = \min(d_{t,k}, 2D_i)$. 
Then, based on the Lagrangian reformulation in \eqref{eq:lagrangian}, the reward at time step $t$ is computed by:
\begin{equation}
r_t = 
\frac{b_t}{N_\mathrm{ts}}
- \lambda_q \sum_{i \in \mathcal{I}} \sum_{k \in \mathcal{K}_i}  (1-\frac{q_{t,k}}{{Q}_i})^+ - \lambda_d \sum_{i \in \mathcal{I}} \sum_{k \in \mathcal{K}_i}  (\frac{\hat{d}_{t,k}}{{D}_i}-1)^+,
\label{eq:reward}
\end{equation}
where $\lambda_q$ and $\lambda_d$ are empirically set.

\item \textbf{Policy $\pi$:} 
The agent's goal is to learn a policy $\pi(\boldsymbol{a}_t|\boldsymbol{s}_t)$ that maximizes the reward function in Eq~\eqref{eq:reward} over time.

\item \textbf{Transition:} 
$P( \boldsymbol{s}_{t+1}|\boldsymbol{s}_t,\boldsymbol{a}_t )$ defines how the system state evolves from the current state $\boldsymbol{s}_t$ to the next state $\boldsymbol{s}_{t+1}$ after taking action $\boldsymbol{a}_t$.

\end{itemize}

%% file: 4_Methodology.tex
\subsection{Overview}





\begin{figure}
    \centering
    \includegraphics[width=\linewidth]{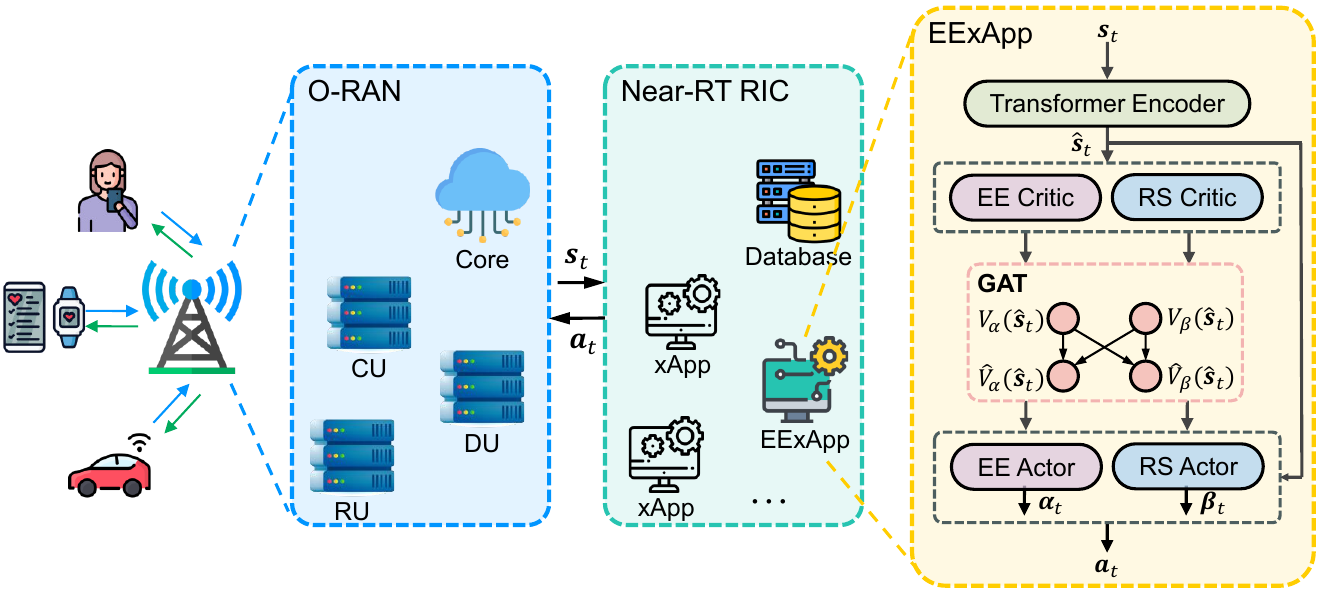} 
    \caption{The architecture of \pname in O-RAN system. The \pname is deployed on the Near-RT RIC, interacting with the RAN (CU and DU) to perform closed-loop control. Internally, the framework utilizes a Transformer encoder to extract latent features from real-time network states $\boldsymbol{s}_t$. A dual-actor-dual-critic structure is employed to decouple the optimization of energy efficiency (EE) and resource slicing (RS), coordinated by a bipartite GAT that aggregates critic values to balance conflicting objectives in the joint action $\boldsymbol{a}_t$.}
    \label{fig:model}
\end{figure}

To solve the above MDP problem, we propose \pname, a dual-actor-dual-critic DRL framework based on \textit{Proximal Policy Optimization (PPO)}. 
Fig.~\ref{fig:model} shows the architecture of \pname.
It deploys two actor networks: \eeactor and \rsactor.
\eeactor is responsible for generating discrete \textit{sleep scheduling} decisions, and \rsactor outputs continuous \textit{resource slicing} decisions. 
The two actors use separate updating strategies, enabling \pname to prioritize distinct optimization objectives:
\eeactor focuses on energy efficiency, while \rsactor learns to maximize QoS satisfaction.

To support these objectives, \pname incorporates two dedicated critics (\texttt{EE Critic} and \texttt{RS Critic} in Fig.~\ref{fig:model}) that independently evaluate energy efficiency and QoS satisfaction. 
\pname is a modular DRL for joint energy efficiency and QoS optimization in O‑RAN control. Built on the model‑free PPO algorithm, it separates decision‑making into two coordinated policies in \eeactor and \rsactor. A lightweight Transformer Encoder processes variable‑length UE observations into fixed‑size representations, enabling adaptation to dynamic network conditions. A bipartite GAT mechanism fuses critic evaluations to capture the coupling between energy‑saving and QoS objectives, supporting stable and effective joint optimization in highly variable wireless environments.

The design of \pname faces two challenges. 
One key challenge is that the input state to \pname is dynamic in size and composition, due to fluctuations in the number of UEs connected to the gNB and the set of active slices over time. Specifically, the number of UEs changes rapidly in real-world deployments as devices arrive, leave, or hand over between cells. This presents a difficulty because the PPO models, which underlie \pname, require a fixed-size input data dimensionality for both the actor and critic networks. To address this challenge, we employ a \textit{Transformer Encoder} to process the UE state information before policy generation. The Transformer Encoder allows \pname to represent a set of UE observations of varying size in a unified, fixed-dimensional latent space.
Details are provided in \S\ref{subsec:encoder}.

The second challenge lies in the modeling of interaction between the two actor-critic pairs for joint optimization. 
In O-RAN control, resource slicing and sleep scheduling target distinct objectives: one for QoS satisfaction and the other for energy efficiency. 
While using separate actor-critic pairs enables specialization, it overlooks their potential interaction. For instance, increasing sleep duration may degrade QoS, while strict QoS enforcement reduces energy-saving opportunities. To model this inter-dependence while maintaining modularity between actor-critic pairs, we introduce a bipartite GAT that connects two critic nodes to two actor nodes in a bipartite graph. The attention mechanism learns how much each critic should influence each actor, allowing each actor to receive a weighted combination of both critic values based on their relevance to its policy update.
Details are provided in \S\ref{subsec:gat}.

\subsection{Transformer Encoder for Dynamic Input Representation} 
\label{subsec:encoder}



At each timestep $t$, \pname acquires a set of real-time observations $\boldsymbol{s}_t = [ \boldsymbol{s}_{t,1}, \boldsymbol{s}_{t,2}, \ldots, \boldsymbol{s}_{t,K} ] \in \mathbb{R}^{K \times F}$ from the DU through the E2 interface, where $K$ denotes the number of active UEs. Each $\boldsymbol{s}_{t,k} \in \mathbb{R}^F$ consists of $F$ KPI features of the network. The raw features $[\boldsymbol{s}_{t,k}]$ are first projected into a latent embedding space:
\begin{equation}
    \boldsymbol{e}_{t,k} = \boldsymbol{W}_e\boldsymbol{s}_{t,k} + \boldsymbol{b}_e, \quad \boldsymbol{e}_{t,k} \in \mathbb{R}^d, \quad k\in \mathcal{K},
\end{equation}
where $d$ is the embedding dimension, and $\boldsymbol{W}_e \in \mathbb{R}^{d \times F}$ and $\boldsymbol{b}_e \in \mathbb{R}^d$ are learnable parameters. 
The sequence of projected embeddings $\boldsymbol{e}_{t} = [ \boldsymbol{e}_{t,1}, \boldsymbol{e}_{t,2}, \dots, \boldsymbol{e}_{t,K} ] \in \mathbb{R}^{K \times d}$ is then processed by a lightweight \textit{Transformer Encoder} tailored for the low-latency and near-RT inference in wireless networks. Specifically, our Encoder comprises 2 self-attention layers, with a reduced embedding dimension $d=64$ and 4 attention heads. The feedforward network hidden size is reduced to $2d=128$, significantly reducing parameters and inference complexity while retaining Transformer’s key properties: contextual embedding of UE states and scalability to dynamic input sizes. This yields a sequence of contextually enriched UE embeddings:
\begin{equation}
    \boldsymbol{h}_t = [ \boldsymbol{h}_{t,1}, \boldsymbol{h}_{t,2}, \ldots, \boldsymbol{h}_{t,K} ], \quad \boldsymbol{h}_{t,k} \in \mathbb{R}^d,
\end{equation}
The output at this stage still reflects the dynamic size of the input (i.e., it depends on the size of $K$).
To satisfy the fixed-size input requirement of the policy optimization, we apply a mean pooling operation over all UE embeddings to construct a fixed-size global context vector:
\begin{equation}
    \hat{\boldsymbol{s}}_t = \frac{1}{K}\sum_{k=1}^{K}\boldsymbol{h}_{t,k},
\end{equation}
Consequently, the resulting global state $\hat{\boldsymbol{s}}_t$ is a fixed-size vector of length $d$.

\subsection{Dual-Actor-Critic Architecture}
From a systems‑level perspective, maintaining two specialized policies instead of a single unified one enhances modularity and adaptability. While a unified policy risks over‑generalization and suboptimal performance across diverse conditions, distinct policies can be tailored to capture regime‑specific dynamics, enabling targeted optimization and greater robustness.

The encoded state $\mathbf{\hat{s}}_t$ is then processed in parallel by actor and critic networks. The discrete \eeactor is responsible for sampling a sleep time decision 
$\boldsymbol{\alpha}_t \sim \pi_\alpha(\boldsymbol{\alpha}_t | \mathbf{\hat{s}}_t)$ 
from a categorical policy over $\boldsymbol{\alpha}_t = (a_t, b_t, c_t)$. Simultaneously, the continuous \rsactor samples a {slicing decision} 
$\boldsymbol{\beta}_t \sim \pi_\beta(\boldsymbol{\beta}_t | \mathbf{\hat{s}}_t)$ from a Gaussian policy. 
After applying the joint action $\boldsymbol{a}_t = [ \boldsymbol{\alpha}_t, \boldsymbol{\beta}_t ]$ to the O-RAN system, the environment transitions to the next observation $\mathbf{s}_{t+1}$ and returns the reward $r_t$. The global reward is decomposed into two components, $r_{t, \alpha}$ and $r_{t, \beta}$, targeting energy efficiency and QoS satisfaction, respectively:

{\small
\begin{equation}
\begin{cases}
r_{t, \alpha} = \dfrac{b_t}{N_\mathrm{ts}}, \\[6pt]
r_{t, \beta} = - \lambda_q \sum\limits_{i \in \mathcal{I}} \sum\limits_{k \in \mathcal{K}_i} (1 - \dfrac{q_{t,k}}{Q_i} )^+ 
- \lambda_d \sum\limits_{i \in \mathcal{I}} \sum\limits_{k \in \mathcal{K}_i} ( \dfrac{\hat{d}_{t,k}}{D_i} - 1 )^+.
\end{cases}
\end{equation}
}

Parallel to the actors, the critics estimate the expected returns. The discrete critic $V_\alpha(\mathbf{\hat{s}}_t)$ estimates the value for the sleep action, while the continuous critic $V_\beta(\mathbf{\hat{s}}_t)$ estimates the value for the slicing action. These raw value estimates are further processed to form \textit{aggregated} critic estimates, $\hat{V}_\alpha(\mathbf{\hat{s}}_{t})$ and $\hat{V}_\beta(\mathbf{\hat{s}}_{t})$, which incorporate a weighted combination of both critics via a bipartite GAT (detailed in \S\ref{subsec:gat}).

To compute the training targets, we adopt the Generalized Advantage Estimation (GAE) method to balance bias and variance over finite-horizon trajectories \cite{schulman2017proximal}. At each timestep $t$, the GAE is calculated for each actor using its corresponding reward and the \textit{aggregated} value estimate. For \eeactor (and similarly for the \rsactor), the Temporal Difference (TD) error $\delta_{t,\alpha}$ is defined as:
\begin{equation}\delta_{t,\alpha} = r_{t,\alpha} + \gamma \hat{V}_\alpha(\mathbf{\hat{s}}_{t+1}) - \hat{V}_\alpha(\mathbf{\hat{s}}_t),
\end{equation}where $\gamma \in [0,1]$ is the discount factor. The advantage estimate $A_{t,\alpha}$ is then obtained by exponentially weighting future TD errors with the GAE parameter $\lambda \in [0,1]$:
\begin{equation}A_{t,\alpha} = \sum_{l=0}^{T-t-1}(\gamma \lambda)^{l} \delta_{t+l,\alpha},
\end{equation}
where $T$ denotes the trajectory length.

The actor network, parameterized by $\theta_{\alpha}$, is updated by maximizing the Clip objective function:

{\small
\begin{equation}
\mathcal{L}(\theta_\alpha) = \mathbb{E}_t \left[ \min \left( \rho_t(\theta_\alpha) A_{t,\alpha}, \ \text{clip}(\rho_t(\theta_\alpha), 1-\epsilon, 1+\epsilon ) A_{t,\alpha} \right) \right],
\label{eq:actor_loss}
\end{equation}
}

\!\!\!\!\!\!where $\rho_t(\theta_\alpha) = \frac{\pi_{\theta_\alpha}(\boldsymbol{\alpha}_t|\mathbf{\hat{s}}_t)}{\pi_{\theta_{\text{old}}}(\boldsymbol{\alpha}_t|\mathbf{\hat{s}}_t)}$ is the probability ratio, and $\epsilon \in [0.1, 0.2]$ is the clipping hyperparameter.

The critic network, parameterized by $\phi_\alpha$, minimizes the error between the predicted value $V_\alpha(\mathbf{\hat{s}}_t)$ and the target return $R_{t,\alpha}$. Consistent with GAE, the target return is defined as the sum of the current value estimate and the computed advantage:
\begin{equation}
R_{t,\alpha} = \hat{V}_\alpha(\mathbf{\hat{s}}_t) + A_{t,\alpha}.
\end{equation}
To enhance the robustness, we employ the Huber loss:
\begin{equation}
\mathcal{L}(\phi_\alpha) = \mathbb{E}_t \left[ f{_\zeta}( V_\alpha(\mathbf{\hat{s}}_t) - R_{t,\alpha} ) \right],
\label{eq:critic_loss}
\end{equation}
where $f_{\zeta}(\cdot)$ is the Huber loss function with threshold $\zeta$:
\begin{equation}
f_{\zeta}(x) =
\begin{cases}
\frac{1}{2}x^2 & \text{if } |x| \leq \zeta,\\
\zeta\left( |x| - \frac{1}{2}\zeta \right) & \text{otherwise}.
\end{cases}
\end{equation}
The resource slicing actor-critic pair ($\theta_\beta, \phi_\beta$) follows the identical update procedure using $r_{t,\beta}$ and $\hat{V}_\beta$.

\subsection{GAT for Critic Aggregation} 
\label{subsec:gat}

In O-RAN control, resource slicing and sleep scheduling target distinct objectives: one for QoS satisfaction and the other for energy efficiency. Employing separate actor-critic pairs allows for specialized learning, but it overlooks the interplay between these objectives. For example, extending sleep duration may compromise QoS, whereas rigid QoS requirements can limit opportunities for energy savings.
To model this interplay while maintaining modularity between actor-critic pairs, we introduce a GAT that connects two critic nodes to two actor nodes within a bipartite graph structure. The attention mechanism dynamically learns the degree to which each critic should influence each actor, allowing each actor to incorporate a weighted combination of critic values relevant to its policy update.

To model the relationship between the separate critic values 
(i.e., ${V}_\alpha(\mathbf{\hat{s}}_t)$ and ${V}_\beta(\mathbf{\hat{s}}_t)$) 
and the aggregated critic values 
(i.e., $\hat{V}_\alpha(\mathbf{\hat{s}}_t)$ and $\hat{V}_\beta(\mathbf{\hat{s}}_t)$), 
we create a lightweight bipartite graph $G = (\mathcal{S}, \mathcal{T}, \mathcal{E})$, 
where 
$\mathcal{S} =  \{S_\alpha, S_\beta\}$ represents the set of source nodes, 
$\mathcal{T} =  \{T_\alpha, T_\beta\}$ represents the set of target nodes,
and
$\mathcal{E} = \mathcal{S} \times \mathcal{T}$ denotes the edges. 
For the two source nodes, the node features are defined as the raw critic estimates: $V_\alpha(\mathbf{\hat{s}}_t)$ for $S_\alpha$ and $V_\beta(\mathbf{\hat{s}}_t)$ for $S_\beta$. 
For the two target nodes, we utilize their corresponding raw estimates as the initial query features: $V_\alpha(\mathbf{\hat{s}}_t)$ for $T_\alpha$ and $V_\beta(\mathbf{\hat{s}}_t)$ for $T_\beta$.  
Leveraging the bipartite GAT mechanism, we update the feature of target nodes by:

\begin{equation}
    \hat{V}_j(\mathbf{\hat{s}}_t) =  \sum_{i \in \{\alpha, \beta\}} \gamma_{ij} w_s V_i(\mathbf{\hat{s}}_t) , 
    \quad
    j \in \{\alpha, \beta\},
\end{equation}
where
$\gamma_{ij}$ is attention score given by
$\gamma_{ij}=\frac{\exp (e_{ij})}{\sum_{i' \in \{\alpha, \beta\}} \exp (e_{i'j})}$ 
with
$e_{ij} = \operatorname{LeakyReLU}\big([w_s V_i(\mathbf{\hat{s}}_t), w_t \hat{V}_j(\mathbf{\hat{s}}_t)] \mathbf{p}^\top\big)$
for 
$i, j \in \{\alpha, \beta\}$.
Here, $w_s \in \mathbb{R}$ and $w_t \in \mathbb{R}$ are learnable parameters, 
$\mathbf{p}$ is learnable parameters as well. 

The features of the target nodes, i.e., 
$\hat{V}_\alpha(\mathbf{\hat{s}}_t)$ and $\hat{V}_\beta(\mathbf{\hat{s}}_t)$, represent the aggregated critic value functions, which are used to update the policy in each timestep as described above.

\subsection{Training Logic}

\begin{algorithm}[!t]
\caption{EExApp training process.} 
\label{algorithm}
\begin{algorithmic}[1]
\Statex {Input:} Learning rates $\eta_\theta$, $\eta_\phi$, discount factor $\gamma$, GAE parameter $\lambda$

\Statex {Initialize:} $\theta_\alpha$, $\theta_\beta$ for actors, $\phi_\alpha$, $\phi_\beta$ for critics, bipartite graph $G = (\mathcal{S}, \mathcal{T}, \mathcal{E})$

\For{$t = 1, \ldots, T$}
    \State Observe the network state $\boldsymbol{s}_t$
    
    \State Encode the state $\mathbf{\hat{s}}_{t}$ using Transformer

    \State Execute actions $\boldsymbol{a}_t = [ \boldsymbol{\alpha}_t, \boldsymbol{\beta}_t ]$ sampled from $\pi_{\theta}(\boldsymbol{a}_t|\mathbf{\hat{s}}_{t})$ and store the transition $(\mathbf{\hat{s}}_{t}, \boldsymbol{a}_t, r_{t,\alpha}, r_{t,\beta}, \mathbf{s}_{t+1})$

    \State Calculate raw critic values $V_\alpha(\mathbf{\hat{s}}_t), V_\beta(\mathbf{\hat{s}}_t)$ and aggregate them into $\hat{V}_\alpha(\mathbf{\hat{s}}_t), \hat{V}_\beta(\mathbf{\hat{s}}_t)$ via GAT

    \State Compute the advantage estimate $A_{t,\alpha}, A_{t,\beta}$ using GAE based on collected trajectory

    \State Compute the probability ratio $\rho_t(\theta_\alpha), \rho_t(\theta_\beta)$

    \State Optimize the clipped objective function $\mathcal{L}(\theta)$ for each actor: $\theta \leftarrow \theta - {\eta}_{\theta} \nabla_{\theta} \mathcal{L} (\theta)$

    \State Update the value function (critic) using Huber loss $\mathcal{L}(\phi)$: $\phi \leftarrow \phi - \eta_{\phi} \nabla _{\phi}\mathcal{L}(\phi)$

\EndFor
\end{algorithmic}
\end{algorithm}

The training process of \pname is summarized in Alg.~\ref{algorithm}. Each training iteration begins with the collection of interaction trajectories between \pname and the O-RAN environment. At each timestep $t$, the \pname receives a sequence of KPI observations $\boldsymbol{s}_t$, which are encoded into a fixed-dimensional state representation $\boldsymbol{\hat{s}}_t$ via a Transformer-based encoder. This latent state then serves as the input to all actor and critic networks. For each state $\boldsymbol{\hat{s}}_t$, two actor-critic pairs operate in parallel: 
the discrete  \eeactor samples a sleep decision $ \boldsymbol{\alpha}_t\sim \pi_\alpha(\boldsymbol{\alpha}_t |\boldsymbol{\hat{s}}_t ) $  from a categorical policy, while the continuous \rsactor samples a slicing decision $\boldsymbol{\beta}_t \sim \pi_\beta(\boldsymbol{\beta}_t|\boldsymbol{\hat{s}}_t )$ from a Gaussian policy, 
At the same time, a critic $V_\alpha({\boldsymbol{\hat{s}}}_t)$ estimates the return from the sleep action $\boldsymbol{\alpha}_t$, while another critic $V_\beta({\boldsymbol{\hat{s}}}_t)$ evaluates the long-term reward from the slicing action $\boldsymbol{\beta}_t$. 
These value estimates are passed to a bipartite GAT, which models the interdependence between the two control tasks. The GAT produces aggregated critic values $\hat{V}_\alpha({\boldsymbol{\hat{s}}}_t)$ 
and 
$\hat{V}_\beta({\boldsymbol{\hat{s}}}_t)$, which are used in subsequent advantage estimation and policy updating.


%% file: 5_Experiments.tex
In this section, we present a series of experiments to evaluate the effectiveness of the proposed \pname. Specifically, we aim to address the following research questions:
\begin{itemize}
    \item \textbf{Q1 (\S \ref{B})}: How does \pname perform in the realistic O-RAN deployment?
    \item \textbf{Q2 (\S \ref{B})}: Does the dual-actor-dual-critic architecture outperform its single-actor-single-critic counterpart? 
    \item \textbf{Q3 (\S \ref{C})}: What is the individual impact of the Transformer encoder and the GAT-based aggregator on the overall system?
    \item \textbf{Q4 (\S \ref{D})}: How does \pname compare against the SOTA solutions?
\end{itemize}

\subsection{Implementation and Experimental Setup} 
\label{A}

\begin{figure}[!t]
    \centering
    \includegraphics[width=\linewidth]{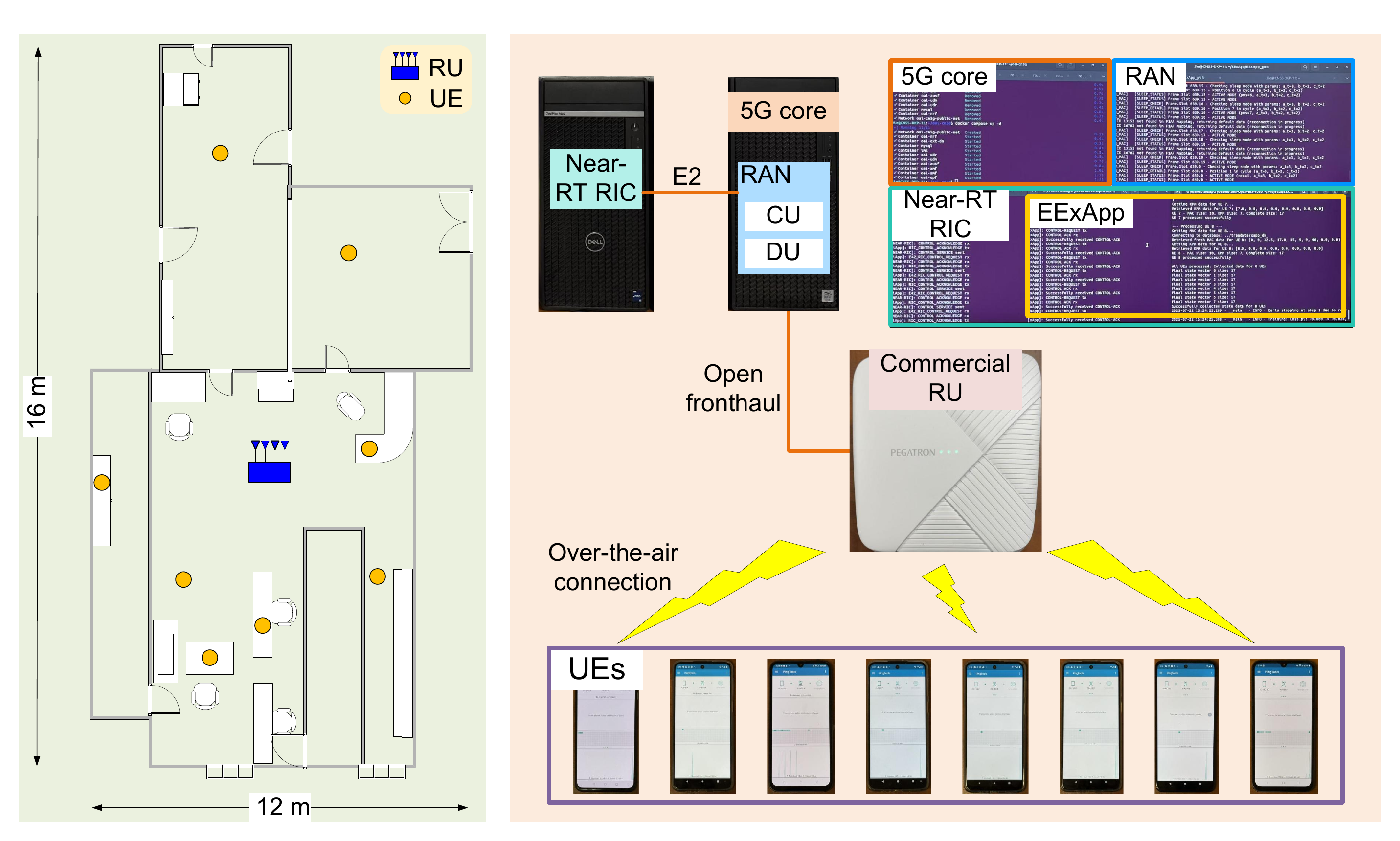}\vspace{-0.05in}
    \caption{Our O-RAN system for experimental evaluation. The left subfigure shows the floor plan of our system deployment, and the right subfigure shows the network elements and architecture.}
    \label{fig:testbed}
\end{figure}

We established an end-to-end experimental testbed to evaluate \pname in a realistic indoor environment. As illustrated in Fig.~\ref{fig:testbed}, our system integrates a 5G Core, RAN components (CU and DU), a commercial RU, a Near-RT RIC, and eight commercial smartphones serving as UEs. As shown in Fig.~\ref{fig:testbed} (Left), the experiment is conducted in a laboratory floor plan where the RU is fixed at a specific location, and the UEs are distributed across different positions to capture diverse channel conditions and spatial dynamics. Fig.~\ref{fig:testbed} (Right) further details the network topology and hardware connections among these components. Specific hardware specifications and software versions are listed in Table~\ref{tab:testbed}. 

On the software side, we implemented the O-RAN system using open-source repositories from OAI for both the 5G core and RAN components \cite{oai_nrsatutorial2025, oai5g2025}. To enable fine-grained sleep scheduling and resource slicing, we modified the underlying OAI 5G codebase: specifically, we updated \path{openair2/E2AP/RAN_FUNCTION} for slicing service modeling and \path{openair2/LAYER2/NR_MAC_gNB} for real-time gNB transmission control. On the hardware side, the Pegatron PR1450 serves as the commercial RU, configured with 4×4 MIMO antennas and operating in the n78 band (TDD mode) with a 30 kHz subcarrier spacing. \pname is deployed on the Near-RT RIC using the FlexRIC platform \cite{flexric2025}, interacting with the RAN via the E2AP (v2.03) and KPM (v2.03) service models. The UEs consist of heterogeneous smartphones running various Android versions.

For evaluation, we use the reward function defined in \S\ref{PF}, along with the QoS violation ratio, which quantifies service degradation due to unmet QoS requirements. The single-actor-single-critic (SASC) approach serves as the primary baseline to validate the benefits of our decoupled architecture.

\begin{table}[!t]
\renewcommand{\arraystretch}{1.0}
\caption{Key components of our testbed.}
\label{tab:testbed}
\footnotesize
\centering
\scriptsize
\begin{tabular}{lll}
\toprule
\textbf{Unit} & \textbf{Hardware} & \textbf{Software} \\
\midrule
5G Core  & Intel Core i7-12700 &  OAI CN5G \cite{oai_nrsatutorial2025}  \\
RAN & Intel Core i7-12700  &  OAI 5G \cite{oai5g2025} \\
SDR & Pegatron PR1450 & 1.0.3.1p4 \\
Near-RT RIC & Intel Core i7-10700 & \makecell[l]{FlexRIC with E2AP\\v2.03 and KPM v2.03 \cite{flexric2025}}\\
UEs & \makecell[l]{OnePlus Nord AC2003, Motorola \\ G54 5G, Samsung Galaxy A15 5G} & Android 11, 13, 14, 15 \\
\bottomrule
\end{tabular}
\end{table}

\subsection{Case Studies under Diverse Network Conditions} \label{B}

To evaluate the robustness of \pname under diverse network conditions, we conduct extensive experiments across varying traffic loads and slicing configurations against the SASC baseline. We define three per-UE target traffic levels generated via \texttt{iPerf} (UDP): light (0.1–1 Mbps), medium (1–5 Mbps), and heavy (5–10 Mbps). Furthermore, we evaluate scenarios with 2, 4, and 8 slices, with the 8 UEs distributed evenly across them.

\begin{figure}[!t]
\centering
\includegraphics[width=\linewidth]{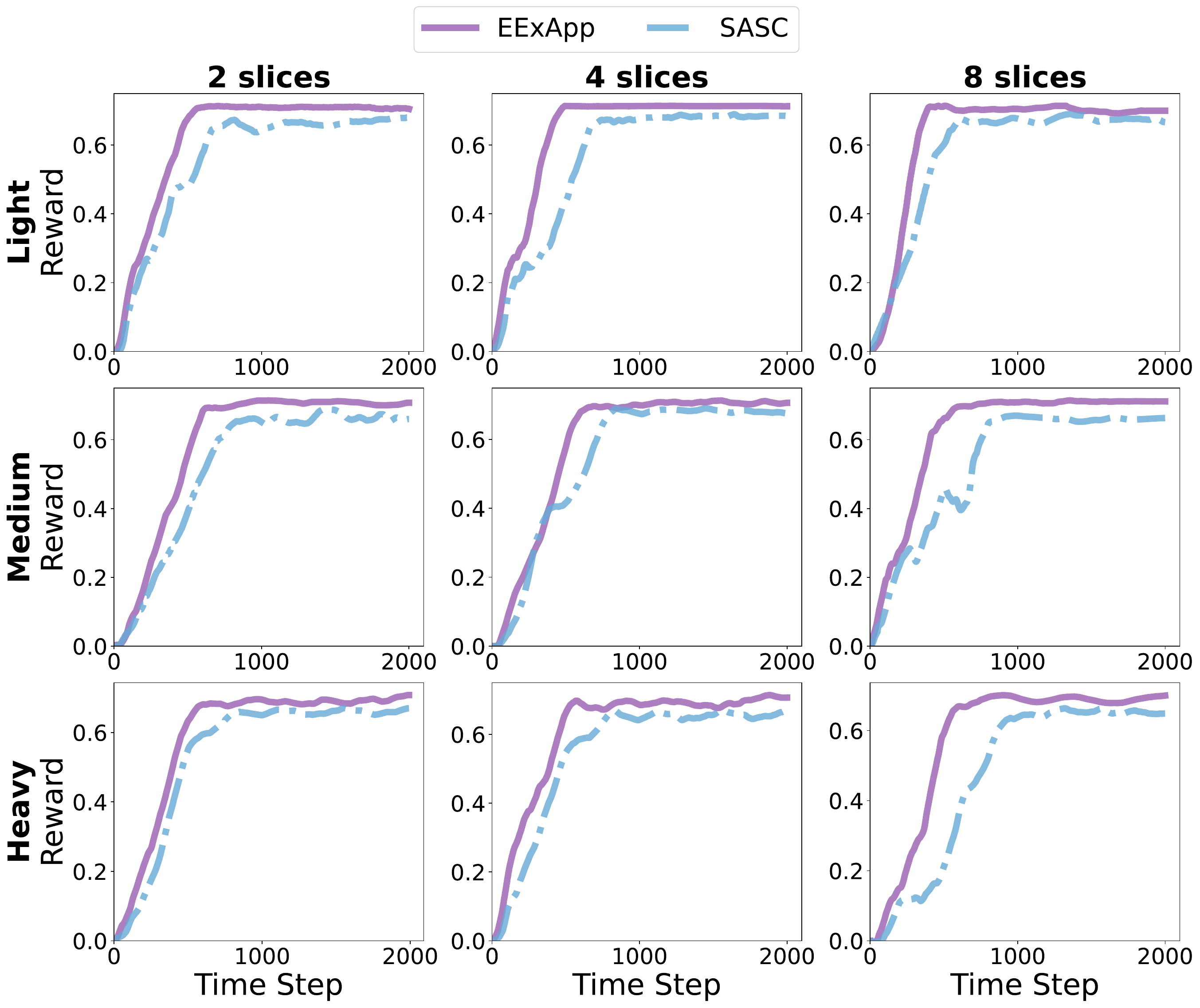}\vspace{-0.05in}
\caption{Convergence performance comparison of \pname and SASC under light, medium, and heavy traffic conditions.}
\label{fig:reward}
\end{figure}

Fig.~\ref{fig:reward} illustrates the convergence performance. \pname consistently stabilizes around 500 timesteps, significantly faster than SASC, which typically converges after 750 steps. As traffic load and slice count increase, both methods exhibit more significant reward fluctuations. This phenomenon occurs because increasing the number of slices leads to resource fragmentation. With fewer resources dedicated to each slice, the statistical multiplexing gain is diminished, limiting scheduling flexibility. In such cases, even minor under-provisioning can cause QoS violations, which negatively impact the reward.

Moreover, under heavy traffic, the system operates closer to or beyond its capacity, intensifying resource contention and QoS violations (detailed in \S\ref{D}). Concurrently, high buffer occupancy constrains RU sleep opportunities, thereby reducing energy efficiency. These factors collectively degrade reward performance, corroborating our theoretical analysis.

While both models achieve comparable asymptotic rewards under light loads, \pname exhibits superior convergence speed. In complex scenarios involving high loads or dense slicing, \pname outperforms SASC in both convergence rate and final reward. This demonstrates the sample efficiency of the dual-actor-critic architecture. By decoupling optimization objectives, \pname mitigates gradient conflicts inherent in competing tasks. Conversely, SASC suffers from slower adaptation and heightened volatility due to the challenge of balancing trade-offs within a monolithic policy. A detailed comparison of QoS violations is provided in \S\ref{D}.

\subsection{Ablation Studies} 
\label{C}

In this subsection, we conduct the ablation studies to evaluate the individual contributions of the key components within \pname. Specifically, we investigate the impact of two critical modules: the Transformer Encoder and the GAT aggregator. We derive three variants of \pname by selectively removing or replacing these components:
\begin{itemize}
    \item \textbf{w/o Trans:} The Transformer Encoder is replaced with a 2-layer MLP, removing the capability to model temporal dependencies and contextual UE states.
    \item \textbf{w/o GAT:} The GAT module used for coordinating dual critics is removed, forcing the actor updates to rely solely on independent critic feedback.
    \item \textbf{w/o Both:} Both Transformer and GAT are removed.
\end{itemize}

\begin{figure}
    \centering
    \includegraphics[width=0.5\linewidth]{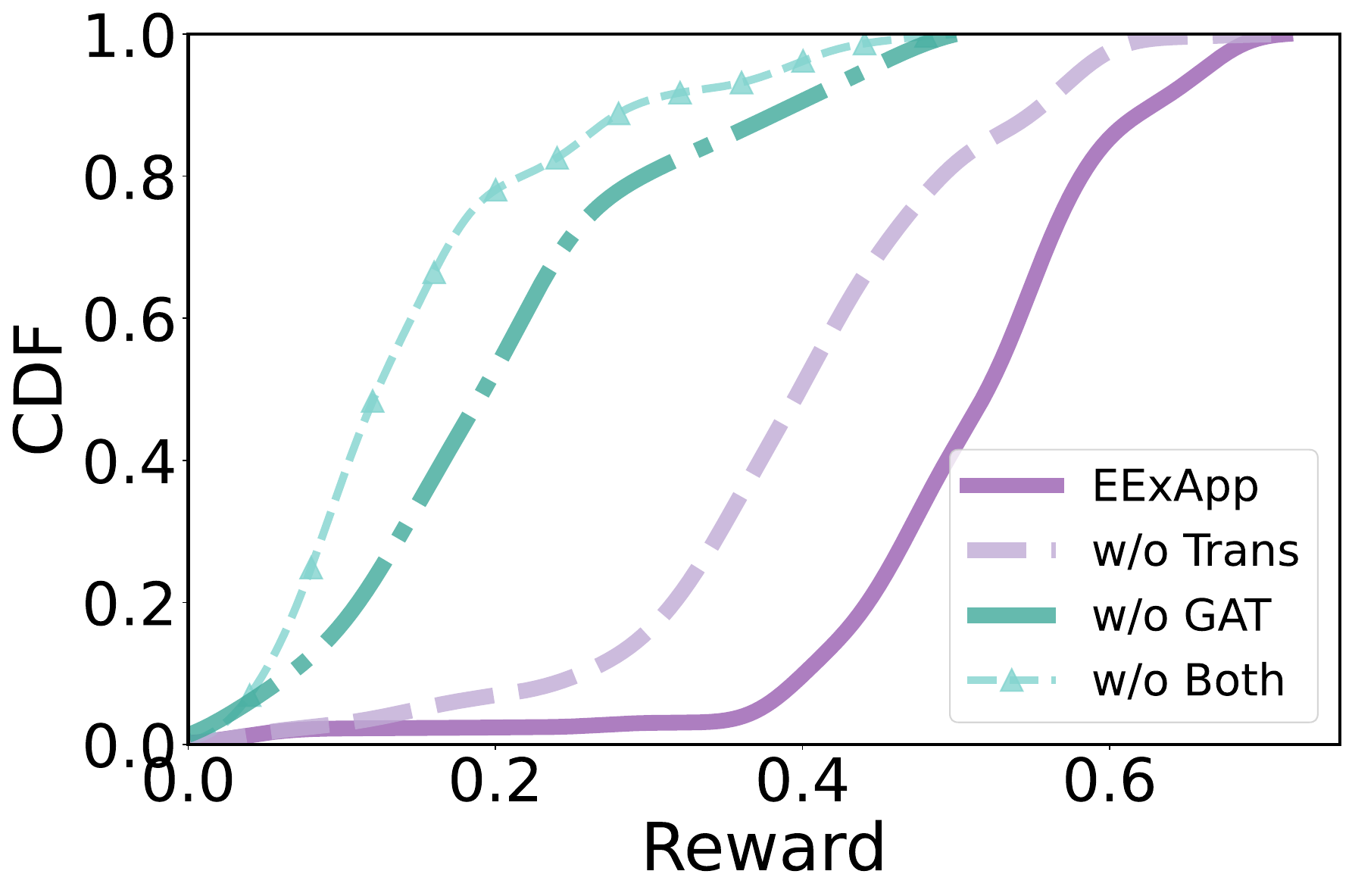}\vspace{-0.0in}
    \caption{The reward CDF of ablation studies.}
    \label{fig:ab_reward}
\end{figure}
\begin{figure}[!t]
    \centering
    \includegraphics[width=\linewidth]{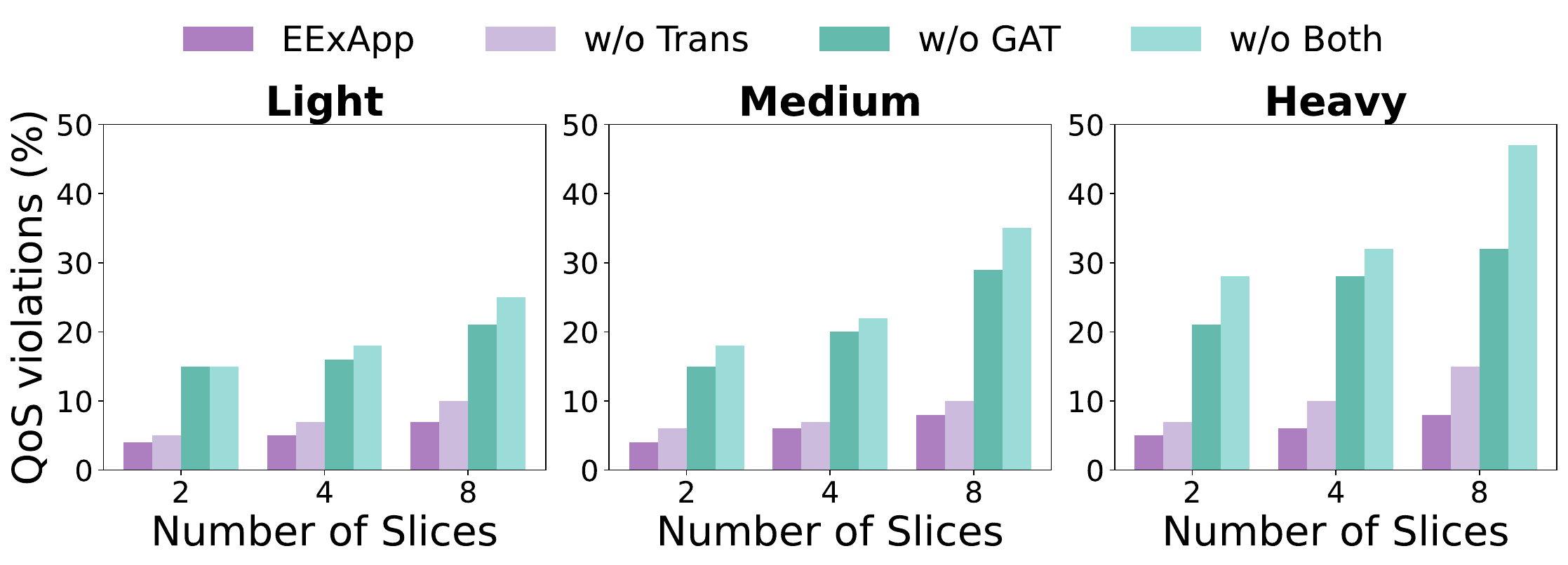}\vspace{-0.05in}
    \caption{The QoS violation comparison of ablation studies.}\vspace{-0.15in}
    \label{fig:ab_qos}
\end{figure}

All models were evaluated across nine diverse load and slicing scenarios described in \S\ref{B}. The cumulative distribution function (CDF) of the rewards aggregated across all scenarios is shown in Fig.~\ref{fig:ab_reward}.

The results demonstrate that both the Transformer encoder and the GAT aggregator play critical roles in the overall performance of \pname. The full \pname model consistently achieves higher rewards than any of its ablated variants. Removing the Transformer module (\textbf{w/o Trans}) leads to significant performance degradation, highlighting its necessity in capturing temporal dependencies and contextual user relationships. Similarly, removing the GAT module (\textbf{w/o GAT}) results in a notable performance drop, demonstrating the value of coordinated dual-critic feedback. The combined removal of both components (\textbf{w/o Both}) yields the worst performance, approaching the lower bound of achievable rewards.

The QoS violation results in Fig.~\ref{fig:ab_qos} further corroborate these observations. As traffic load increases, all model variants exhibit rising QoS violations, particularly as the number of slices grows. Nevertheless, \pname consistently maintains superior robustness with the lowest violation rates under all conditions. Notably, the performance gap between \pname and its ablated versions widens significantly under high traffic loads and complex slicing. This indicates that the Transformer and GAT modules are essential for maintaining system stability in saturated network environments.



\subsection{Comprehensive Performance Comparison}
\label{D}

To evaluate the performance of \pname against SOTA radio energy-saving solutions, we compare it with the following algorithms:

\begin{itemize}
\item \textbf{Kairos} \cite{lozano2025kairos}: An xApp-based single-actor–multi-critic DRL architecture. A centralized actor determines the delay allowance, while multiple distributional critics independently estimate energy savings and QoS compliance for each network slice.
 
\item \textbf{O-RAN DRL} \cite{O-RAN_DRL}: A DRL-based framework that optimizes the trade-off between UE throughput and energy consumption using various PPO and DQN models. For a fair comparison, we adopt the PPO-1 variant, identified as the top-performing model in \cite{O-RAN_DRL}.

\item \textbf{SASC}: The single-actor–single-critic version of \pname, as detailed in \S\ref{A}.

\end{itemize}

\begin{figure}
    \centering
    \includegraphics[width=0.55\linewidth]{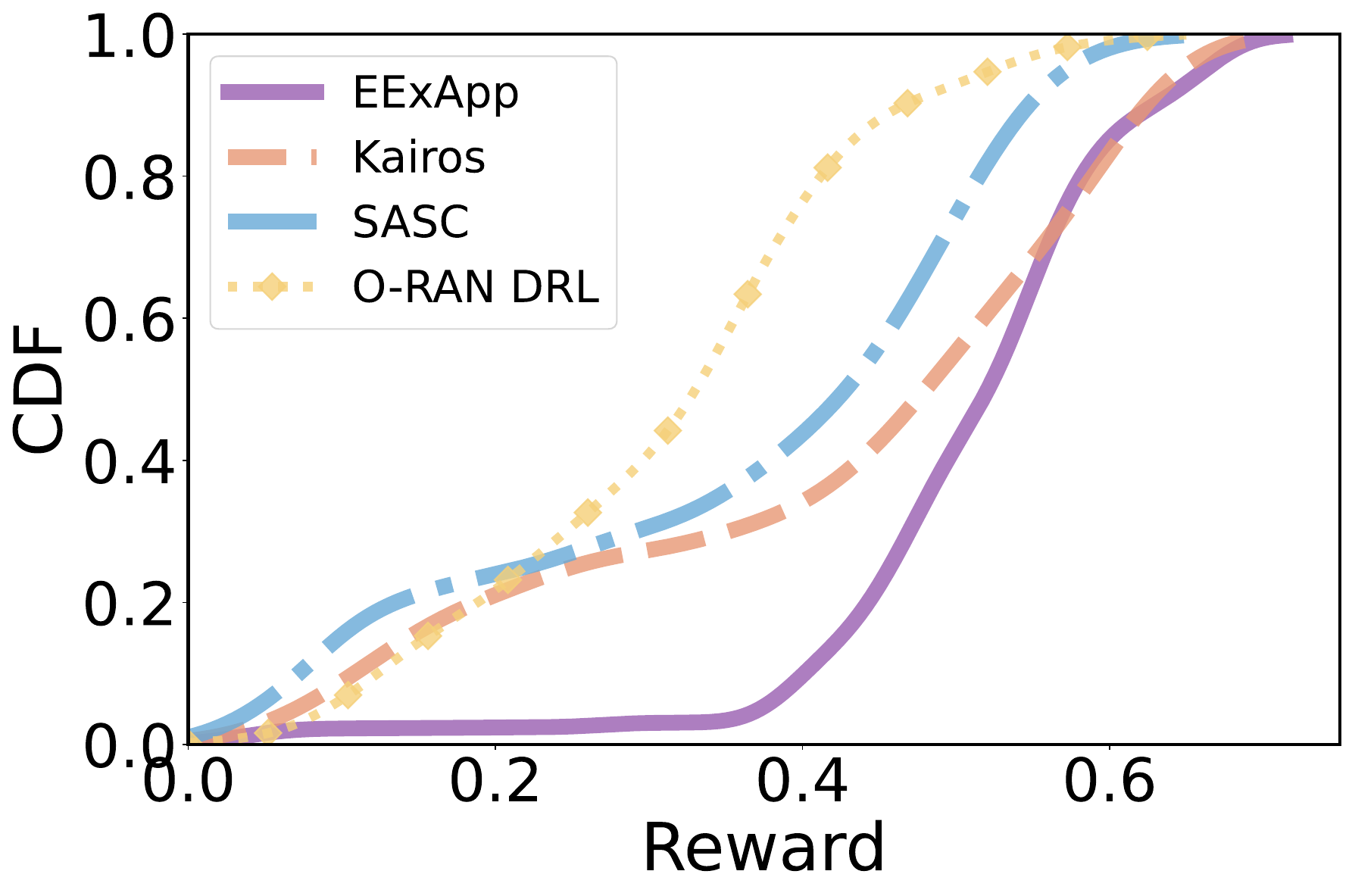}\vspace{-0.05in}
    \caption{Reward comparison of \pname and SOTA baselines.}
    \label{fig:comp_reward}
\end{figure}

We adapted these baselines to align with our problem formulation and the reward function defined in \S\ref{PF}. All experiments are conducted under the nine hybrid load and slicing scenarios outlined in \S\ref{B}.

The results, summarized in Fig.~\ref{fig:comp_reward} and Fig.~\ref{fig:comp_qos}, demonstrate that \pname consistently outperforms the baselines in balancing energy efficiency and QoS. Regarding reward distribution (Fig.~\ref{fig:comp_reward}), \pname achieves superior cumulative rewards compared to Kairos, SASC, and O-RAN DRL. In terms of QoS violations (Fig.~\ref{fig:comp_qos}), \pname maintains consistently low violation rates across all load conditions and slice counts, particularly under medium and heavy traffic.

Kairos achieves the lowest QoS violation rates overall, owing to its distributional critics that explicitly monitor QoS compliance. However, \pname surpasses Kairos in overall reward by identifying a more effective trade-off boundary. While Kairos adopts a conservative policy that restricts sleep opportunities to strictly minimize violations, \pname leverages the dual-actor architecture to safely extend sleep durations, thereby maximizing energy savings without engaging in excessive QoS violations. In contrast, both SASC and O-RAN DRL underperform, exhibiting lower rewards and higher violation rates. Their performance degrades under complex network conditions, such as high traffic loads or large slice counts, indicating limited robustness and adaptability.

%% file: 2_Related_Work.tex
\textbf{Radio Energy Saving in 5G:} 
The energy-saving topic in RAN has been extensively investigated, which can be summarized into four main strategies: i) \textit{time-domain strategies} that schedule idle periods via sleep modes, such as discontinuous transmission (DTX) and discontinuous reception (DRX) \cite{de2025network, vatanian2024energy, oikonomakou2023power}; ii) \textit{frequency-domain strategies} that reduce bandwidth usage or deactivate carriers, such as carrier aggregation \cite{khoramnejad2022delay, wei2023carrier} and carrier shutdown \cite{de2023modeling}; iii) \textit{power-domain strategies} that optimize power amplifier efficiency, such as transmit power adaptation \cite{lu2024resource, wang2025batteryless}; iv) \textit{spatial-domain strategies} that employ dynamic antenna and transceiver activation/deactivation such as antenna element adaptation \cite{van2021exploiting, xu2023integrated, lu2023sic, lu2023uav}.

\begin{figure}[!t]
    \centering
    \includegraphics[width=\linewidth]{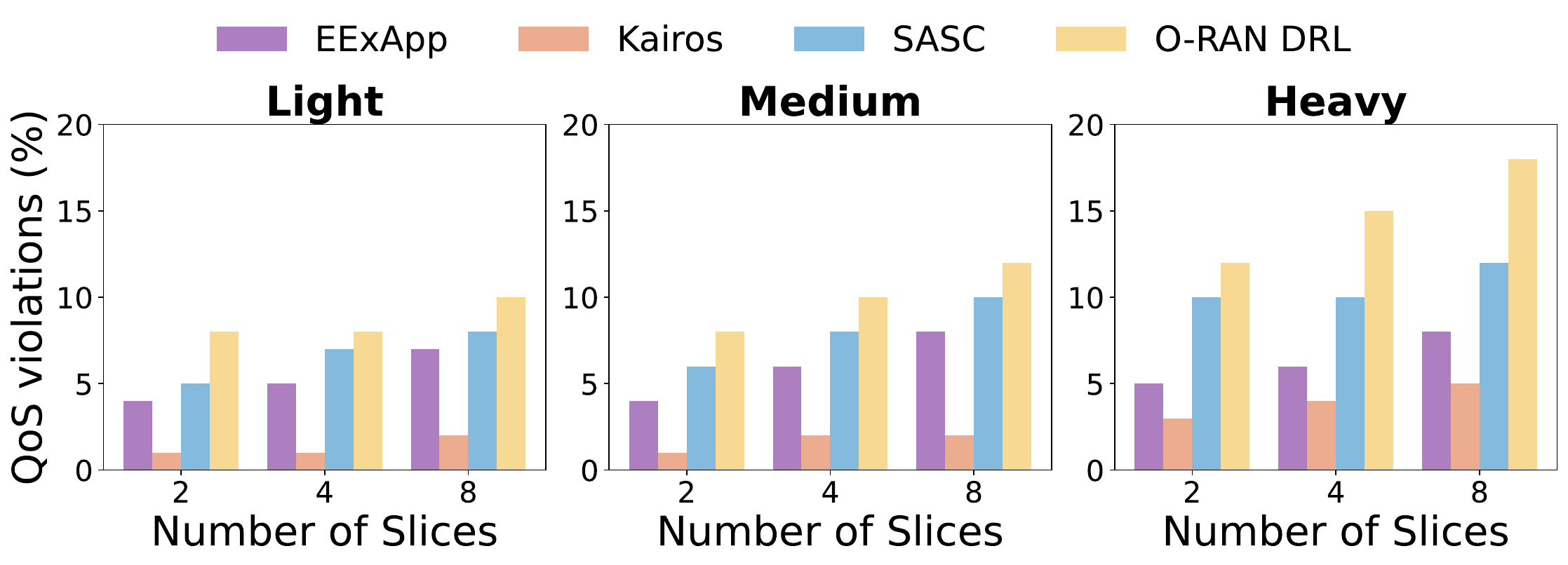}\vspace{-0.05in}
    \caption{QoS violation comparison of \pname and SOTA baselines.} 
    \label{fig:comp_qos}
\end{figure}

However, the efforts to integrate energy-saving techniques into O-RAN systems are still in their early stages. 
Existing efforts vary widely in their design scope. For instance, works like RLDFS \cite{RLDFS} and CFMM \cite{CFMM} focus on function splitting, while BSvBS \cite{BSvBS} and EEDRA \cite{EEDRA} address resource allocation. Other studies, such as ScalO-RAN \cite{ScalO-RAN}, investigate compute scaling. Despite this diversity, many approaches remain non-learning-based or rely on static heuristics. While effective in specific settings, such approaches often lack adaptability to dynamic traffic patterns and struggle to optimize long-term trade-offs, such as balancing energy versus QoS.



\textbf{RL for Energy Optimization in O-RAN:} 
RL-based approaches have recently gained attention within the O-RAN context. 
Early studies, including the aforementioned RLDFS \cite{RLDFS}, employed basic Q-learning and Sarsa to minimize energy consumption, while ES-xApp \cite{ES-xApp} applied DQN to switch off radio components. More recent advancements, such as Kairos \cite{lozano2025kairos} and O-RAN DRL \cite{O-RAN_DRL}, have extended the scope to advanced sleep mode (ASM) and joint optimization of throughput, energy, and mobility, leveraging Actor–Critic or PPO strategies. These works demonstrate RL's potential in managing complex trade-offs in O-RAN energy optimization. 

Building on this progress, \pname advances RL-enabled energy saving in O-RAN through three key contributions: (i) Joint optimization of RU sleeping scheduling and DU resource slicing; (ii) Specialized dual-actor-critic framework to handle conflicting objectives under dynamic traffic; and (iii) Real-world validation via over-the-air experiments, moving beyond purely simulation-based evaluations.


%% file: 6_Conclusion.tex
This paper proposed \pname, a DRL-based xApp designed for 5G O-RAN systems to jointly optimize RU sleep scheduling and DU resource slicing. By leveraging a dual-actor–dual-critic PPO framework, \pname effectively balances the trade-off between energy efficiency and QoS compliance, decoupling the optimization targets to specialized agents. The Transformer-based encoder enables scalable processing of dynamic UE populations, while the bipartite GAT module coordinates the distinct critics to support robust and adaptive policy updates. Extensive over-the-air experiments on a real-world testbed demonstrate that \pname achieves substantial reductions in RU energy consumption while maintaining QoS, consistently outperforming SOTA baselines under diverse network conditions. 